\documentclass[12pt,preprint]{aastex}

\shorttitle{The Radio Spectrum of TVLM513-46546}
\shortauthors{Osten et al.}

\begin{document}

\title{ The Radio Spectrum of TVLM513-46546: Constraints on the Coronal
Properties of a Late M Dwarf
}
\author{Rachel A. Osten\altaffilmark{1}}
\affil{National Radio Astronomy Observatory, 520 Edgemont Road, Charlottesville, VA 22903; 
rosten@nrao.edu}
\altaffiltext{1}{Jansky Postdoctoral Fellow}

\author{Suzanne L. Hawley}
\affil{Astronomy Department, Box 351580, University of Washington, Seattle WA 98195; slh@astro.washington.edu}

\author{Timothy S. Bastian}
\affil{NRAO, Charlottesville, VA 22903; tbastian@nrao.edu}

\author{I. Neill Reid}
\affil{Space Telescope Science Institue,  3700 San Martin Drive, Baltimore MD
21218; inr@stsci.edu}

\begin{abstract}
We explore the radio emission
from the M9 dwarf, TVLM513-46546, at multiple radio frequencies,
determining the flux spectrum of persistent radio emission,
as well as 
constraining the levels of circular polarization. 
Detections at both 3.6 and 6 cm provide spectral index measurement
$\alpha$ (where S$_{\nu} \propto \nu^{\alpha}$) of $-0.4\pm0.1$.
A detection at 20 cm suggests that the spectral peak is between 1.4 
and 5 GHz.  
The most stringent upper limits on circular polarization are at 3.6 and 6 cm,
with $V/I <$15\%.
These characteristics agree well with those of 
typical parameters for early to mid M dwarfs, 
confirming that magnetic activity is present 
at levels comparable with those extrapolated from earlier
M dwarfs.
We apply analytic models to investigate the coronal properties under simple
assumptions of dipole magnetic field geometry and radially varying
nonthermal electron density distributions.
Requiring the spectrum to be optically thin
at frequencies higher than 5 GHz 
and reproducing the observed 3.6 cm fluxes
constrains the magnetic field at the base to be less than about 500 G.
There is no statistically significant periodicity in the
3.6 cm light curve, but it is consistent with low-level variability.
\end{abstract}

\keywords{stars: activity, stars: coronae, stars: late-type, 
radio continuum }

\section{Introduction }
The detection of magnetic activity signatures in very late-type dwarfs
(late M$\rightarrow$L) is puzzling, as the fraction of objects showing H$\alpha$
in emission,
the traditional activity signature, declines precipitously past type M8 \citep{west2004}. 
The rotation-activity correlation found in earlier type stars (F to early M) 
appears to break down: most objects M9 and cooler have rapid rotation, but
weak or undetectable H$\alpha$ emission \citep{mohantybasri2003}. 
Models suggest \citep{mohantyetal2002} 
that the highly neutral atmospheres of these ultracool objects should be incapable of
sustaining the kinds of magnetic stresses which pervade the atmospheres of 
solar-like stars (G, K, early M) and participating in the non-radiative heating of chromospheric
and coronal plasma.
And yet,
transition region, X-ray, and radio persistent emission is clearly detected
in some objects \citep{hjk2003,flemingetal2003,berger2002}, along with flare-like variability \citep{rutledge2000,liebert2003}.
Transition region
and X-ray observations detect persistent fluxes at levels compatible with scaling laws
(based on H$\alpha$) extrapolated from earlier type M dwarfs.  The detection of 
both persistent and flaring radio emission from a handful of late M and L objects,
reported in \citet{bergeretal2001} and \citet{berger2002}, suggests that 
nonthermal radio emission
is also present at these late spectral types.  The interpretation of the radio 
emission, however, is complicated by the fact that many potential mechanisms may be
at work.  In an effort to better understand the radio emission from late M dwarfs, we
undertook a deep pointing of an object previously detected at a single radio frequency.

TVLM513-46546 (=2MASS J15010818+2250020) is a young disk M9 dwarf, located
 at a distance of 10.5 pc.  It was first detected in \citet{tinney1993}.
It has  H$\alpha$ emission which is variable \citep[equivalent width measurements of 1.8
 \AA\ and 3.5 \AA;][]{mohantybasri2003,reidetal2002}, and 
rapid rotation (v$\sin i\sim$60 km s$^{-1}$) corresponding to a minimum rotation 
period of 2.4 hours (see \S 5).
There is no evidence of a physical companion between
0.1'' and 5'' \citep{closeetal2003}.  Luminosity and effective temperature estimates
from \citet{leggetetal2001} are $\log (L_{\rm bol}/L_{\odot})\sim$-3.65 and 
T$_{\rm eff}\sim$2200 K.  
The lack of lithium in the spectrum 
constrains it to be more massive than 0.06 M$_{\odot}$ and thus likely
a very low mass star rather
than a brown dwarf.
TVLM513-46546 (hereafter, TVLM513) was detected in the radio
at a wavelength of 3.6 cm by \citet{berger2002}, who gave an average flux density of 
308$\pm$16 $\mu$Jy, with evidence of variability (peak flux of 1100 $\mu$Jy).

\section{Data Reduction}
Observations were made on 2004 January 24 with the NRAO Very Large Array (VLA)\footnote{The 
NRAO is a facility of the National Science Foundation operated under 
cooperative agreement by Associated Universities, Inc} when the array was being 
reconfigured from B to C configuration.  
The observations spanned $\sim$11.3 hrs, using receivers at 20, 6, and 3.6 cm.
The flux density calibrator was 3C286, with assumed flux densities at 
20, 6, and 3.6 cm
of 14.5, 7.4, and 5.2 Jy, respectively.  The phase calibrator at all frequencies 
was 1513+236,
with flux densities of 1.7, 0.8, and 0.5 Jy at 20, 6 and 3.6 cm, respectively.
Data reduction was performed in AIPS.
The observing strategy involved time sharing between frequency bands, with approximately 10 minutes 
on source per frequency; we reduced each frequency separately, first imaging
the field of view to determine if the object was detected, then removing 
all background sources from the visibility dataset and re-imaging the field.
A two-dimensional Gaussian was fit to the source, deriving intensity and
source location.  For frequencies at which a positive detection was made,
the discrete Fourier transform of the visibilities was made as a function of time
using the AIPS task DFTPL to examine variability.  Images in Stokes V were also
made to search for any evidence of circularly polarized flux.  

\section{Analysis}
\subsection{Flux and Polarization Spectra}
TVLM513 was strongly detected at 6 and 3.6 cm, with a marginal detection (5.6$\sigma$)
at 20 cm.
Table~\ref{tbl:fluxes} describes the parameters of the radio source at each frequency
for the time-integrated data.
The derived spectral index between 6 and 3.6 cm is $-0.4\pm0.1$, and between 20 and
6 cm it is 0.1$\pm$0.2.  The upper limits (3$\sigma$) on the amount of circularly polarized flux at
3.6 and 6 cm lead to constraints on the circular polarization V/I 
of $<$ 15\%.  
The persistent nature of the radio flux as well as the lack of circular polarization signal
points to a uniform coverage of the stellar disk with magnetic fields, either large
or small-scale. 

We computed the 
distribution of 6--3.6 cm spectral indices from
a small sample of radio observations of dMe and dKe stars, compiled by \citet{gb1996}.
Although there is a fair amount of scatter (the slope errors range
from 0.05--0.5), the distribution can be fit well by a Gaussian with FWHM=1.5, centered at
$-0.41$, which agrees with the measured value for TVLM513.
The upper limits on circular polarization
are also consistent with the general trends of quiescent radio emission from dMe stars
\citep{whiteetal1989}.

Estimation of the brightness temperature requires separating the observed source flux density
from the source size.  We can place conservative limits by assuming the source extent to be
the stellar diameter; at 10.5 pc, the fluxes listed in Table~\ref{tbl:fluxes}
imply brightness temperatures of a few times 10$^{9}$ K.  This is consistent with
gyrosynchrotron radiation.


\subsection{Variability}
\begin{figure}[!b]
\begin{center}
\includegraphics[scale=0.4,angle=90]{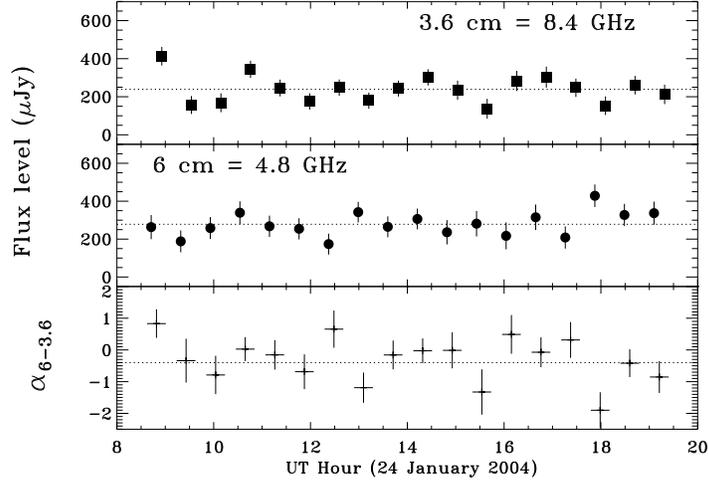}
\figcaption[]{Variation of 3.6 and 6 cm fluxes during the $\sim$10
hour observation.  Each time bin corresponds to an average over each scan,
approximately 10 minutes in length.  The dotted lines indicate the average
flux density during the observation. 
Bottom panel shows variation of 3.6--6 cm spectral
index during the observation; dotted line indicates spectral index determined
from entire observation. \label{fig:vary}}
\end{center}
\end{figure}

Figure~\ref{fig:vary} plots the variation of TVLM513 at the two frequencies
in which it was strongly detected during the course of the observation.  
During the observations of TVLM513 reported in \citet{berger2002} a 
flare (factor of 4 increase in flux density) over $\approx$ 25 minutes was 
recorded.  That radio flare was accompanied by large values
of circular polarization (reaching $\sim$70\%).  In contrast, our
observations appear to be characterized by quiescent conditions.

A Lomb normalized periodogram of the 3.6 cm light curve reveals 
peaks at frequencies corresponding to periods 2.1 and 2.8 hr, 
although the false-alarm probability 
is significant (0.22 and 0.54 for 2.1 and 2.8 hr periods, respectively).
There are no strong periods in the 6 cm light curve.
We examined the periods in the 3.6 cm light curve by imaging
the target in each $\approx$ 10 minute scan, and fitting a two-dimensional
Gaussian to the point source.  The variation of the peak intensity found this
way matches those determined from the DFTPL program, and the same periodicities
are produced from a power spectrum of the light curve.
We would expect a minimum rotation period for TVLM513 of
2.4 hours, given the $v\sin i$ of 60 km s$^{-1}$ and a stellar radius 
of 0.12R$_{\odot}$ \citep[based on direct mass and radius determination for
OGLE-TR-122b of M$=$0.092$\pm$0.009 M$_{\odot}$, 
R$=$0.120$^{+0.024}_{-0.013}$R$_{\odot}$
by radial velocity variations and
transit depths,][]{pont2005}. 


We performed
$\chi^{2}$ tests to determine whether the data were consistent with no variability,
by comparing the fluxes with the average flux density from the observation.
At 3.6 cm, the average flux is 239$\pm$73$\mu$Jy, with 
$\chi^{2}$ fit to the average returning $\chi^{2}_{\nu}$=2.3 (17 dof), indicating
agreement at the 0.2\% level for a constant source.  
At 6 cm, the average flux is 278$\pm$63$\mu$Jy, and a $\chi^{2}$ fit to the
average yields $\chi^{2}_{\nu}$=1.2 (17 dof), 
indicating agreement at the 25\% level
for a constant source.  
We note that the time sampling, 10 minutes in each frequency band every 36 minutes, is relatively coarse, and this together with the low amplitude of the variation may 
affect the statistical significance.  
Thus we can conclude that
there is marginal evidence for variations of the signal 
from TVLM513 at 3.6 cm, but not 6 cm.


Previous observations of TVLM513 with the VLA lasted 6200 seconds 
\citep[23 September 2001, reported in][]{berger2002} and 1350 seconds,
respectively (unpublished observations of 15 February 1994).  Neither had a
sufficient time baseline to identify variability on timescales $\geq$ 2 hours; 
it is unlikely that any periodicities could be seen in those data.
Adding the current observations to this time baseline
reduces the flare duty cycle reported in \citet{berger2002}, 
based on one flare lasting $\sim$1500 s, to $\sim$3\%. 

\section{Coronal Models}
The brightness temperatures inferred in \S 3.1 are too large for 
thermal bremsstrahlung emission
to be the cause.  
Particle acceleration, leading to gyrosynchrotron
emission, is the more probable source.
The spectral shape implies that 
frequencies greater than 5 GHz
are optically thin, and   
the observed 5--8 GHz spectral index is compatible with optically
thin emission from a relatively hard energy distribution with power law index $\delta \sim$2 or 3.
The radio characteristics 
of TVLM513 are very similar to those of active dMe stars, in terms of spectral index
and constraints on polarization levels. 
We investigate simple models for the coronal properties, 
under the assumption that the 
emission mechanism is gyrosynchrotron radiation from a population of mildly relativistic 
electrons: first, for a homogeneous source, and second, for a dipole magnetic field and radially
varying nonthermal electron density.  These two approaches give different, but consistent,
results for the magnetic field strengths in the TVLM513's corona and the
total number density of accelerated particles.

The turnover in spectral index
between 1.4 and 4.8 GHz suggests that the spectral maximum is in this range.  
The spectral turnover could be due to a transition
between optically thick/thin regimes; 
if it is, the peak frequency of gyrosynchrotron emission depends 
on the magnetic field strength in the radio-emitting source, the
source size, and index of the power-law distribution of nonthermal electrons. 
The strongest dependence of these parameters
is from
the magnetic field strength. 
For the simplest case of a homogeneous radio-emitting source, we constrain 
the magnetic field strength
using equation (39) of \citet{dulk1985}, 
setting $\delta$=2(3), source size $\sim$R$_{\star}$--2R$_{\star}$,
and the total number density of nonthermal electrons above a cutoff value of 10 keV
between 10$^{5}$--10$^{6}$ cm$^{-3}$. Field strengths of a few hundred Gauss are needed in
a homogeneous radio-emitting source to produce a spectral turnover below 5 GHz
due to optical depth effects.

\citet{whiteetal1989} investigated simple non-thermal
models for the quiescent radio emission from dMe flare stars,
using an analytic representation of the magnetic field 
and spatial inhomogeneities in the nonthermal electron density distribution. 
The magnetic field is assumed to
be dipolar ($B(r)\propto r^{-n}$, where $n=3$).  
The distribution of electrons with 
energy is parameterized as a power-law with index $\delta$,  and is assumed to vary radially
with distance as $N=N'(E_{0}) (r/r_{0})^{-m}$, where $N'(E_{0})$ is the electron density distribution.
Combining these assumptions with the analytic expressions for absorption and emission coefficients
of \citet{dulkmarsh1982},
expressions for the optically thick and thin radiative fluxes can be obtained\footnote{There is an error in \citet{whiteetal1989} equation 7:  a plus sign in the third line of the formula
should be a multiplication symbol.}.  
\citet{whiteetal1989} considered a global dipole, which they termed a ``bare'' dipole,
and a ``buried dipole'', in which the depth below the surface at which the dipole is buried
is much less than a stellar radius.
\citet{whiteetal1989} compared their analytic expressions with more accurate numerical
calculations and conclude that 
both are within a factor of two of the numerical calculations.
We use the 
analytic expressions to obtain a general description
of coronal conditions which could be applicable to the case of TVLM513.

\begin{figure}[h]
\begin{center}
\includegraphics[scale=0.6]{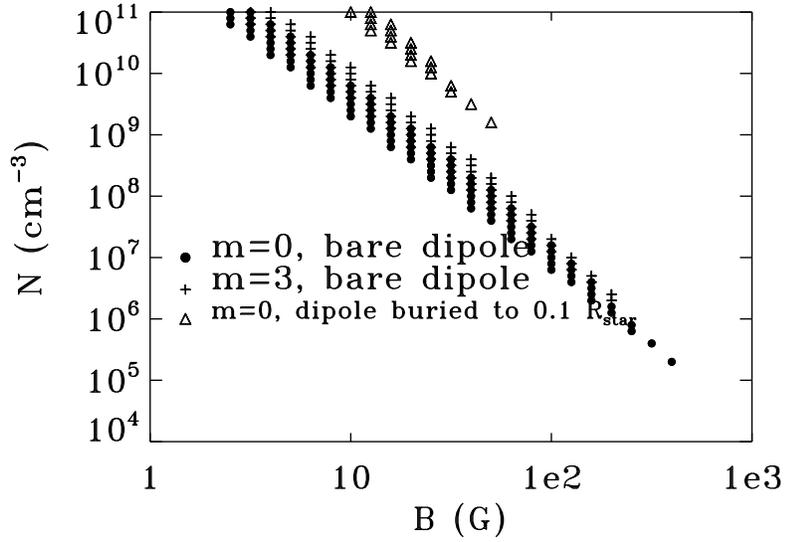}
\figcaption[]{Constraints on dipole magnetic field strength and nonthermal electron density
in the corona of TVLM513, under different assumptions.  
Here, $m$ is the index of the
power-law radial dependence of the nonthermal electron density, and we consider
dipoles whose scale equals the stellar radius (``bare dipole'') and those with scale lengths
less than the stellar radius (``buried dipole'').
Allowed models are those which predict between 0.5 and 2 times the detected flux at 3.6 cm,
and are optically thin at frequencies higher than 5 GHz.
\label{fig:constraints}}
\end{center}
\end{figure}

In our case, we calculate the expected fluxes at the distance of TVLM513 (10.5 pc),
specifying the 
frequency (1.4, 5, and 8 GHz), constraining $\delta$ to be 3, and assuming only dipolar
field configurations.  For a harder electron energy spectrum ($\delta =2$)
the total integrated energy in fast electrons does not converge.
This leaves four
free parameters:  the power-law index describing the radial dependence of the number density ($m$),
the base magnetic field strength ($B_{0}$), the total surface number density ($N_{0}$), 
and the scale length of the dipole ($=x\times R_{\star}$, where $x$ is $\leq 1$).
The $m=0$ case corresponds to a nonthermal electron distribution independent of radius,
appropriate for an isotropic pitch angle distribution.  The other extreme, $m=n=3$, corresponds
to the radial dependence of the electrons being the same as that of the dipolar magnetic field;
this is the situation that would be obtained by conserving both particle and magnetic flux in
the case of open field lines.
We require
that the turnover frequency be $<$5 GHz, based on the observational
constraints. 
Figure~\ref{fig:constraints} shows the range of parameter space in base magnetic field
strength and total nonthermal number density, for 
different combinations of $m$ and bare/buried dipoles which predict between 0.5 and 2 times the detected flux at 3.6 cm.
For $\delta=3$, the predicted flux varies as $S \propto N B^{2.5}$.  The turnover
frequency likewise depends more strongly on the base magnetic field strength than 
the total number density of accelerated electrons.
This places a constraint on the base magnetic field
strength to be less than $\sim$500 G to satisfy the observed fluxes and spectral distribution.
Requiring the dipole to be buried at depths less than the stellar radius increases the 
number density of accelerated electrons needed to produce the same observed flux.

The radio observations indicate little net polarization, but do not
provide a stringent constraint on the global magnetic geometry:  the upper limits
on polarization could be consistent with small regions of magnetic field permeating
the surface, or a large dipolar field with high inclination so that the net
circular polarization signature is small.  \citet{muteletal1987} noted a relationship 
for RS CVn binary systems (known to have large-scale magnetic fields) in which the 
measured circular polarizations
decrease with increasing inclination. 
The high rotational
velocity 
might also suggest a nearly edge-on viewing
geometry to see the maximum rotational broadening.

\section{Conclusions}
There are various lines of evidence
which suggest that similar activity is being displayed on late M dwarfs as is seen
in the 
earlier M dwarfs:  
radio flares; transition region and X-ray emission
(including flaring at these wavelengths); and sporadic detections of H$\alpha$ emission
and flaring.
To this list we can add quiescent radio emission with typical
spectral indices of earlier M dwarfs and little to no circular polarization. 
Longer term observations of ultracool dwarfs are needed to understand fully
the nature of the radio emission and 
 how commonly it occurs in these objects. 
A more comprehensive view of activity diagnostics
in ultracool dwarfs is also needed to make sense of the broad changes seen in H$\alpha$ emission
trends, and to relate them to the physical changes taking place in these ultracool dwarf
atmospheres.


This paper represents the results of VLA project AO180. RAO acknowledges support from a 
Jansky Postdoctoral Fellowship from the National Radio Astronomy Observatory.
SLH acknowledges support from NSF grant AST02-11559.

\clearpage


\clearpage

\begin{deluxetable}{llllllll}
\tablewidth{0pt}
\tablenum{1}
\tablecolumns{8}
\tablecaption{Radio Parameters \label{tbl:fluxes}}
\tablehead{\colhead{$\lambda$} &\colhead{Freq.}& \colhead{T$_{\rm int}$} & \colhead{Beam} & \colhead{I Flux} & \colhead{V Flux} & \colhead{$\pi_{c}$} 
& \colhead{L$_{\rm rad}$} \\
\colhead{(cm)} &\colhead{(GHz)} &\colhead{(s)} & \colhead{''x''} & \colhead{($\mu$Jy)} & \colhead{($\mu$Jy)}
& \colhead{\%} & \colhead{(erg s$^{-1}$ Hz$^{-1}$)} }
\startdata
3.6 &8.4 &10510 & 2.8$\times$2.2, PA=71 & 228$\pm$11 & $<$33 & $<$15 &3.0$\times$10$^{13}$\\
6 &4.8 &10520 & 4.4$\times$3.2, PA=80 & 284$\pm$13 & $<$42 & $<$15 &3.8$\times$10$^{13}$\\
20 &1.4 & 10580 & 13.3$\times$10.1, PA=24 & 260$\pm$46 & $<$68 & $<$26 &3.5$\times$10$^{13}$\\
\enddata
\end{deluxetable}



\end{document}